\documentclass[a4paper,
               biblatex,     
               ]{jacow}
\makeatletter%
	\ifboolexpr{bool{xetex}}
	 {\renewcommand{\Gin@extensions}{.pdf,%
	                    .png,.jpg,.bmp,.pict,.tif,.psd,.mac,.sga,.tga,.gif,%
	                    .eps,.ps,%
	                    }}{}
\makeatother

\ifboolexpr{bool{xetex} or bool{luatex}} 
 {}                                      
 {\usepackage[utf8]{inputenc}}           

\usepackage[USenglish]{babel}
\usepackage{mathtools}
\usepackage{svg}
\usepackage{makecell}
\ifboolexpr{bool{jacowbiblatex}}%
 {
   \addbibresource{biblio.bib}
 }{}
\begin{document}

\title{Laser Pulse Duration Optimization\\ with Numerical Methods}

\author{Francesco Capuano\textsuperscript{1,}\thanks{\texttt{Francesco.Capuano@studenti.polito.it}}, 
		Davorin Peceli, Bedřic Rus, Alexandr Špaček\\
		ELI, Institute of Physics, Czech Academy of Sciences, Doln\'{i} Bre\v{z}any, Czech Republic\\ 
		Gabriele Tiboni\thanks{\texttt{Gabriele.Tiboni@polito.it}}, Politecnico di Torino, Turin, Italy  \\
		\textsuperscript{1}also Politecnico di Torino, Turin, Italy}
	
\maketitle

\begin{abstract}
In this study we explore the optimization of laser pulse duration to obtain the shortest possible pulse.

We do this by employing a feedback loop between a pulse shaper and pulse duration measurements. 
We apply to this problem several iterative algorithms including gradient descent, Bayesian optimization and genetic algorithms, using a simulation of the actual laser represented via a semi-physical model of the laser based on the process of linear and nonlinear phase accumulation. 
\end{abstract}

\section{Introduction \& Related Works}
\subsection{L1-Allegra System}
The L1 Allegra system is a high power laser system developed at ELI Beamlines in Czech Republic. The system is designed to deliver <20fs pulses with an energy higher than 100 mJ at a repetition rate of 1\,kHz. This system is based on the amplification of frequency-chirped pico-second pulses in an Optical Parametric Chirped Pulse Amplification (OPCPA) chain consisting of seven amplifiers. L1 contains 7 OPCPA stages that are pumped by 5 diode pumped lasers based on commercial Yb-doped thin-disk Regenerative Amplifiers (RA) DIRA 200-1. All of the pump lasers generate pulses at 1030\,nm with 1\,kHz repetition rate and also include stretcher, compressor and an LBO crystals for second harmonic generation (SHG) of pulses at 515\,nm. To achieve high SHG efficiency several features of the pump laser system should be optimized. One of these features is the laser pulse temporal shape. The optimization of temporal pulse shape is accomplished through the manipulation of the pulse spectral phase. 
A standard method for spectral phase representation employs a Taylor expansion of spectral phase, $\varphi(\omega)$, around the central angular frequency $\omega_0$, of the spectral pulse:
\begin{equation}\label{eq:phase_exp}
    \varphi(\omega) = \sum_{i=0}^{\infty} \frac{{\partial^{(i)} \phi}}{\partial \omega^{(i)}} \cdot \big(\omega - \omega_0)^i
\end{equation}
The first two terms of the sum of Eq.~\eqref{eq:phase_exp} represent constant phase and group delay and don’t have an effect on the pulse temporal shape. Third, fourth and fifth term are instead related to dispersion parameters such group delay dispersion ($\text{GDD}  \sim \varphi'$), third order dispersion ($\text{TOD}  \sim \varphi''$) and fourth order dispersion ($\text{FOD}  \sim \varphi'''$) and do have an impact on the temporal profile. In this work, we aim at control these parameters, to which we will collectively refer to as $\psi$.

Since SHG is an intensity-dependent non-linear optical process, the highest conversion efficiency is obtained for those pulses with temporal shape closest to transform-limited pulses. In particular, since the transform-limited (TL) pulse is the pulse with minimum possible pulse duration given the present spectral bandwidth, it is guaranteed to have the highest possible pulse intensity.

Temporal properties of amplified pulses are controlled through manipulation of parameters of the stretcher by \textit{Teraxion} and monitored using \textit{Femtoeasy}'s SH FROG 1030. 
The control of pulse temporal shape is done by manipulating three parameters, $d_2, d_3$ and $d_4$ on the stretcher controls that relate to parameters GDD, TOD and FOD previously introduced through a determinate system of linear equations, depending on the central wavelength (derived from central frequency) of the spectrum considered. 

\subsection{Related Works}
Obtaining short and sharp laser pulses is crucial for running high-efficiency physical systems that would reach really high intensities with limited energy consumption.

Various algorithms were used in ultra-fast laser systems for active feedback optimization of laser parameters or laser-matter interactions~\cite{PaperBaumert, PaperDRL}, but none of these previous approaches took care of optimizing the single pulse shape. 

In this work we compare three different algorithms for the optimization of the temporal profile of laser pulses. These are: Bayesian Optimization, Differential Evolution, and a custom implementation of gradient descent.

\subsection{Semi-Physical Model of L1-Pump-System}

\begin{figure}[t!]
    \centering
    \includegraphics[width=\columnwidth]{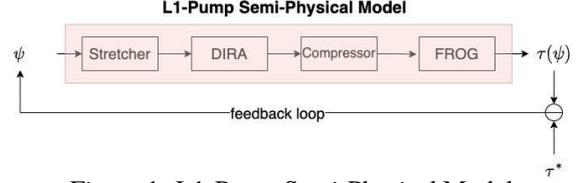}
    \vspace*{-2\baselineskip}%
    \caption{L1-Pump Semi-Physical Model.}
    \label{fig:semi-physical}
    \vspace*{-\baselineskip}%
\end{figure}

Optimization of laser pulse shape is designed through feedback control loop between stretcher and FROG. To test performances of different optimization algorithms we designed a theoretical semi-physical model of the L1 amplifier chain containing stretcher, DIRA, compressor and FROG, as presented in Fig.~\ref{fig:semi-physical}.

Each of the elements in the pump chain contribute to the overall spectral phase.
Furthermore, while stretcher and compressor introduce only linear change to the spectral phase (GDD, TOD and FOD), DIRA also introduces nonlinear effects. The system presented in Fig.~\ref{fig:semi-physical} is parameterized by: \textbf{B-integral} (representing a measure of the accumulation of light's nonlinear phase), estimated to be 2 for our DIRA; \textbf{Compressor parameters}, considered constant in this work and estimated as opposite to the nominal values of stretcher's parameters. We conclude the presentation of our semi-physical model highlighting how DIRA's contribution in terms of linear spectral phase can be neglected.

\section{Pulse Optimization}

Other works did tackle the problem of laser optimization focusing on using Deep Reinforcement Learning to obtain mode-locking inside the laser cavity~\cite{PaperDRL}. 
However, no previous work was done to optimize the actual pulse shape of single pulses to make it as similar as possible to an arbitrary target shape. 

In this work (freely available on GitHub \cite{GitHub}), we focus exactly on this problem. If one indicates with $\tau^*$ such target shape and with $\tau(\psi)$ the temporal shape resulting from the adoption of the control $\psi$ on the system, then it is possible to frame the problem as:
\begin{equation}
    \min_{\psi \in \mathcal X} L\big(\tau^*, \tau(\psi)\big) \ ,
\end{equation}
where $L:\mathbb R^n \mapsto \mathbb R^+$ is a figure of merit which is used to guide the optimization route towards better and better values of $\psi$ and $\mathcal X$ is the region of the parameters space containing physically-feasible configurations.

In this work we adopted as target temporal profile the transform-limited pulse, i.e. the temporal representation of the pulse corresponding to a zero phase imposed on the original spectrum. However, our approach is perfectly robust to different target shapes.

In the case in which the system presented in Fig.~\ref{fig:semi-physical} were not present any non-linear blocks such as regenerative amplifiers, this result can be obtained imposing a control which corresponds to $\psi_{str.} = -\psi_{compr.}$. 
However, since this non-linearity is indeed present in the system (as the beam energy needs to be increased), it is not possible to obtain the transform-limited pulse applying such control. 

\begin{figure}[b!]
    \centering
    \includegraphics[width=\columnwidth]{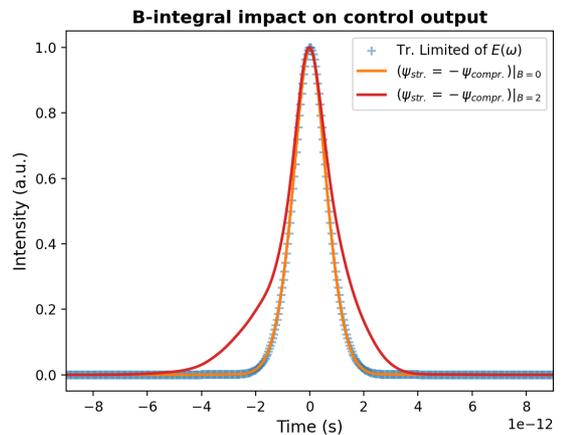}
    \caption{Impact of B-integral and control parameters on output shape.}
    \label{fig:control_pulse}
\end{figure}

This can also be seen in Fig.~\ref{fig:control_pulse}, where it is possible to see that the non linearity present heavily distorts the output of the application of an otherwise optimal control.

\subsection{Loss Function}

We shall first observe that an exact analytical solution to the problem of mixed (linear and non-linear) phase accumulation it is practically impossible to obtain in our case. 

This fact generates the need of adopting a solution that, with an iterative approach, adjusts the control parameters so as to find a region in which the control induces a temporal profile matching the desired one under an acceptable threshold. Such a threshold is computed with respect to an arbitrary, yet very relevant, $L$. 

Our first guess has been the Mean Squared Error (MSE) loss function. 
However, we observed that this loss function is really not informative as a guiding signal for the optimization process. This is due to two main reasons: 
\begin{itemize}
    \item Since the temporal profile is always represented via a 0-1 normalization in Intensity values, absolute differences are also in this range. This implies that differences in the order of magnitude of one tenth -which are very significant in this case- account only for values in the order of magnitude of one cent, when squared (according to MSE formulation). Since the majority of the two curves (the tails of the impulse) numerically assumes very similar values, the MSE's numerator is very likely to be a small number.
    
    \item The algorithm we used to perform Fourier Transformations requires a very large number of points (often ~30k) in order to obtain temporal profiles with acceptable resolutions.
    This essentially means the MSE's denominator would always be a large number, independently on the actual temporal shapes, thus biasing the final value towards small values.
\end{itemize}

In light of this, we used different loss functions for our problem, as MSE would be inevitably biased towards small numbers and, therefore, not a valuable guiding signal in any optimization route.



\subsection{Optimization Algorithms}

To obtain the configuration of parameters which minimizes one of the losses considered, we implemented various optimization algorithms. In particular, we implemented three algorithms: two so-called "black box" algorithms, namely Bayesian Optimization and Differential Evolution, and one "first order" algorithm based on first-order differential information employed in an implementation of the ADAGRAD algorithm~\cite{adagrad}. 

It is important to note that while both the black-box algorithms can, in principle, be applied to the real-world laser in the sake of optimizing its configuration, ADAGRAD cannot. ADAGRAD could be applied if the gradients were to be approximated with finite differences, but this would very much increase the computational effort required.

However, since we plan to collect enough real data to train an Articial Neural Network (NN) to reconstruct the $\psi \mapsto \tau$ mapping (said NN would support automatic differentiation w.r.t. $\psi$), we decided to experiment with this algorithm despite its immediate lack of practical applicability.

\subsubsection{Bayesian Optimization}

Bayesian Optimization is a popular black-box optimization algorithm that probes at random various points inside a feasible region $\mathcal X$, adjusting, according to Bayesian Theorem, a probability distribution on the function fitting best those points. 

This function is referred to as “surrogate function” or, more commonly, “acquisition function”. 
At each iteration, the new point to be probed is selected balancing between exploration of the search space and maximization of this acquisition function.

We want to stress that this algorithm should not be directly applied on the real hardware, as the exploration of the search space causes serious oscillations between subsequent probed points, thus posing the actual dynamical system under considerable stress. 

As a final note, it is important to note that in our implementation we restricted the number of iterations to 150, as we observed performance improvement stagnation over around 120 iterations and obtained the result presented in Fig.~\ref{fig:final_results}. As it is possible to see, the obtained results presents some wings that make it significantly different from the target profile for what concerns pulse mounting and de-mounting. Those wings could probably be mitigated by further exploring the GDD and TOD search space.

Among all the different loss functions tested, the one that gave the best results is the \textbf{sum-L1-Manhattan norm}. We used the default acquisition function-exploration coefficient ($\kappa$) configuration of~\cite{REPO:BOpt} as hyper-parameters.

\subsubsection{Differential Evolution}
Like Bayesian Optimization, Differential Evolution (DE) is a black-box algorithm that explores a unknown landscape in a feasible region $\mathcal X$. 

This algorithm mimics the process of natural selection, initializing a population of candidate solutions that iteratevely gives birth to mixed individuals.

The fitness of the various candidate solutions is typically strictly related to the value of the objective function evaluated at each individual point. 
The algorithm is defined so as to promote some characteristics of the best individual in each generation's offspring, so as to have that as generation go by, the fitness level increases as well as the quality of the candidate points. 

While with this algorithm we observed a significant decrease in oscillations between subsequent points (particularly at the beginning), we must observe that such oscillations are most likely not very dangerous for the actual system, since they are very rare. 

However, this solution may be really slow when applied to the actual machinery. Such a increase in the needed time is mainly due to the fact that, when in the real world, each function evaluation lasts at least as the actual physical process underlying the control application. Since DE is a method which requires a large number of function evaluations, it is clear how this method could turn out to be really slow as minutes may be necessary to complete only one of a multitude of necessary iterations, especially if starting in a region that eventually turns out to be far apart from the optimal point.

Among all the different loss functions tested, the one that gave the best results is a mixture of one which computes the \textbf{MSE only for non-zero values}, (thus discarding the pulses' long tails) and a measure of \textbf{the mismatch in terms of subtended area by controlled and target pulse} (weighted in favour of difference of area, respectively with 0.3-0.7 weights).

\begin{figure}[!b]
    \centering
    \includegraphics[scale=0.475]{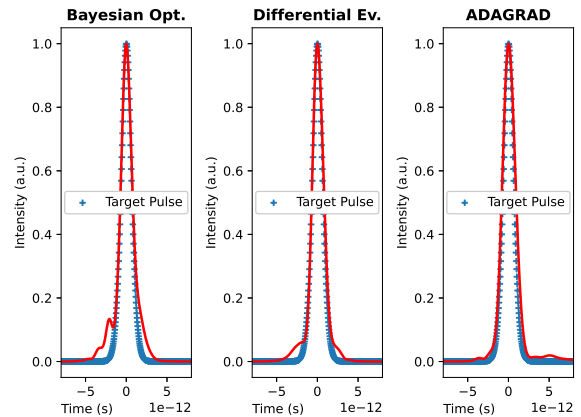}
    \caption{Final results obtained with different algorithms.}
    \label{fig:final_results}
\end{figure}

As a final note, it is important to note that in our implementation we restricted the number of iterations to 150 and the population size to 20. The mutation coefficient, responsible to combine population's individuals has been set to 0.8 whereas the recombination probability between the best candidate and the mutant has been set to 0.7. 
The results obtained with this algorithm turned out to be far better than the ones obtained with Bayesian Optimization and are presented in Fig.~\ref{fig:final_results}. It is possible to see a pedestal which indicates a sub-optimal exploration of the TOD and FOD space, even if the obtained pulse is very much similar to the transform limited for the considered set of parameters.

\subsubsection{ADAGRAD}

Unlike the two other proposed approaches, ADAGRAD assumes the possibility of accessing first-order differential information so as to build an optimization algorithm which ultimately differs from regular steepest descent for its capability of adapting and scaling its step-length with respect to the dimension it is optimizing.

This is particularly interesting for our setting, as the feasible region $\mathcal X$ presents some peculiarities, such as the fact that between the first and the last dimension there are almost than 30~orders of magnitude. Therefore, an adaptive strategy for the learning rate must be put in place. 

ADAGRAD was born as an unconstrained optimization algorithm, and has had a wide adoption in NN-training. To decline this algorithm to this constrained case we employed theoretical results presented in~\cite{BookWright}, such as Sequential Penalty Methods, turning the constrained optimization problem into a constrained one. 

At each iteration, the optimization route is directed in the steepest descent direction using an adaptive learning rate which re-scales an initial learning rate $\eta_0$ using the inner product of all past gradients so far. This helps scaling the step-length using information on past steepest directions which, ultimately, leads to an adaptive learning rate. 
If the new point is outside of the feasible region, then the objective function is penalized with an increasing penalty terms that eventually pushes the optimization route into $\mathcal X$.  

\begin{table}[!t]
	\centering
	\caption{Final Results Obtained}\label{table:results}
	\begin{tabular}{ccc} 
		\toprule
		\textbf{Algorithm} & \bfseries \makecell{Number of \\ func. evaluations} & \bfseries \makecell{Final MSE \\ reached} \\ \midrule
		\textit{Bayesian Opt.} & 150 & \num{2.625e-05} \\
		\midrule
		\textit{Differential Ev.} & \makecell{200 (its) \\ $\times$ 20 (pop-size) \\ = 4000} & \num{1.450e-05} \\
		\midrule
		\textit{ADAGRAD} & 3000 (its) = 3000 & \num{4.899e-06}  \\
		\bottomrule
	\end{tabular}
    \vspace*{-\baselineskip}
\end{table}

This algorithm is definitely the safest in terms of relative dissimilarity between two different candidate points (since this difference can be controlled using $\eta_0$) and the one requiring the smallest number of function evaluations. However, it is practically deployable only in the case in which a NN able to approximate the mapping $\psi \mapsto \tau$ is built. 
Alternatively, this method could be built using finite differences, thus resulting in eliminating the need for the ANN. In this case, the number of function evaluations could be contained to be even one order of magnitude smaller than the number of function evaluations needed in DE, for instance.

Among all the different loss functions tested, the one that gave the best results is the natural logarithm of the sum-L1-Manhattan norm. We used an exponentially increasing penalty term equal to $\frac{1}{10 \cdot (0.9999)^{k}}$ and an initial learning rate of $\eta_0=10$. Moreover, we mapped $\psi$ to its $\mathrm{fs}^2, \mathrm{fs}^3$ and $\mathrm{fs}^4$ representation to improve the numerical stability of the algorithm.

The results obtained with this algorithm are presented in Fig.~\ref{fig:final_results}. As it is possible to see, ADAGRAD practically reproduces the transform-limited pulse, although the right shoulder of the pulse is a bit off. This is most definitely due to an exploration of the GDD dimension.

\subsection{Results}
The algorithms just presented have been all used to optimize the phase so as to obtain the most similar shape to transform-limited. 
It is worth noting that, while we did benchmark our model and observed that it captures some fundamental trend in actual FROG-reconstructed pulse shapes, we did limit the complexity of the model to keep its run-time in bearable time for algorithms employing a typically large number of iterations. 

Our results are presented in terms of obtained shapes in Fig.~\ref{fig:final_results} and, for what concerns the final MSE reached, in Table~\ref{table:results}.

\section{CONCLUSION}

In this work we showed how to effectively use different algorithms to modify the temporal shape of the L1-Allegra high-power laser pulse. The modification of the temporal profile has a wide range of applications, as it is a first step towards to reach the high intensities that users and researchers at ELI Beamlines demand. To obtain such result, we did explore the space of parameters with three different supervised algorithms, as they all needed a target temporal profile to carry out the parameters optimization. 

However, we did observe a high dependency of the results found (respectively, the best configuration of parameters obtained with the different algorithms) on the parametrization of laser environment. This is a crucial problem, as precise knowledge about the laser's parametrization is typically non-obtainable. 

As future work, we plan on expanding the Pulse Optimization process with techniques which are robust to such a problem. In particular, sequential decision making approaches -such as in Reinforcement Learning- may be investigated to achieve adaptive, online control of high-power lasers that further complies with real hardware constraints on machine safety.

\printbibliography

\end{document}